\documentstyle[sprocl,graphicx]{article}

\def\Journal#1#2#3#4{{#1} {\bf #2}, #3 (#4)}

\def\PLB{{\em Phys. Lett.}  B}
\def\PRL{\em Phys. Rev. Lett.}

\def\YF{{\em Yad. Fiz.}}

\def\SJNP{{\em Sov. J. Nucl. Phys.}}

\def\JETP{\em JETP}

\def\JETPL{\em JETP Lett.}

\def\PAN{\em Phys.At.Nucl.}
\def\APP{\em Acta Phys.Polonica}

\def\be{\begin{equation}}
\def\ee{\end{equation}}
\def\bea{\begin{eqnarray}}
\def\eea{\end{eqnarray}}

\begin{document}


\title{ON COOPER PAIRING IN FINITE FERMI SYSTEMS}

\author{A. V. AVDEENKOV}

\address{Institute of Physics and Power Engineering,\\ 249029 Obninsk, Russia;\\
JILA and Department of  Physics, University of Colorado,\\ Boulder, CO, USA
\\E-mail: avdeyenk@murphy.colorado.edu}

\author{S. P. KAMERDZHIEV}

\address{ Institute of Physics and Power Engineering,\\ 249029 Obninsk, Russia
\\E-mail: kamerdzhiev@ippe.obninsk.ru}


\maketitle\abstracts{
In order to analyse the role of the quasiparticle-phonon interaction
 in the origin of nuclear gap, we applied an approach which is similar
to the Eliashberg theory for usual superconductors.
We obtained that the averaged contribution
of the quasiparticle-phonon mechanism to the observed value of
the pairing gap for $^{120}$Sn is 26$\%$ and
the BCS-type mechanism gives 74$\%$ .
Thus,
 pairing is of a mixed nature at least in semi-magic nuclei --
it is due to the quasiparticle-phonon  and BCS  mechanisms,
the first one being mainly
a surface mechanism and the second one mainly a volume mechanism.
The calculations of the strength distribution for  the odd-mass nuclei
$^{119}Sn$ and $^{121}Sn$ have shown that the quasiparticle-phonon
mechanism
mainly improves the description of the observed
spectroscopic factors in these nuclei.
    For the case of nuclei with pairing in
both proton and neutron systems it is necessary to go beyond the
Eliashberg-Migdal approximations and include the vertex correction
graphs in addition to the rainbow ones.
The estimations for spectroscopic factors
performed within a three-level model have
shown that the contribution
of the vertex correction graphs was rather noticeable.
}

\section{Introduction}
As it is known in the microscopic theory of ordinary superconductors,
the BCS model of superconductivity
is the limit case
$g^{2}<<1$ of the Eliashberg theory in which the electron interaction
is determined only by the quasiparticle-phonon coupling g \cite{r1}.
In the microscopic  theory of finite nuclei, the BCS equation for the
gap is used , as a rule, which contains a phenomenologically chosen
pp interaction \cite{r2,r3}. It would be justified if $g^{2}<<1$
 for real nuclei where
g is the correspondingly dimensionless phonon creation
amplitude. However, we have only $g^{2}<1$ in magic \cite{BM}
and in semi-magic \cite{r4} nuclei and the case of strong
coupling $g^{2}>1$ for nuclei with pairing in both nucleon
systems. Therefore , one can expect a contribution of the
 quasiparticle-phonon interaction (QPI) to
the observed nuclear gap.

This question has another important aspect.
As it is known, the most collective low-lying phonons, which make
the largest contribution to the QPI effects in nuclei, are mainly
 surface excitations. Thus the explicit singling out of the
quasiparticle-phonon mechanism of nuclear pairing will make it
possible to answer the old question whether nuclear pairing
is a volume or surface effect.

And last but not the least , a consistent study of the QPI
contribution to pairing should improve the description of
nuclear excitations. At present, this is especially interesting
in connection with the quick development and using the
qualitatively new gamma-ray arrays like EUROBALL and GAMMASPHERE.

The problem of the origin of nuclear pairing has been long discussed
 on a phenomenological level
within the theory of finite Fermi systems using its idea
of the internal (volume) and external (surface) pp interaction
(see \cite{r5}). Microscopically , the question about a contribution
of the QPI  to the gap has been discussed in
\cite{KL} and recently in \cite{Acta,Baran,r6}. In \cite{Baran}
the authors solved a BCS equation with the pp interaction
caused by an exchange of a collective phonon between two quasiparticles.
In \cite{KL,Acta} the authors also considered the "insertion"
graphs which correct single-particle moving.
However, for the nuclear case of $g^{2}>1$ it is necessary to
go beyond the Eliashberg-Migdal approximation which complicates
the problem  strongly.

In this article we calculate the contribution of the QPI
to the observed gap within the approximation of $g^{2}<1$
for the nucleus $^{120}Sn$ taking into account both types
of graphs and also consider the case
of strong coupling $g^{2}>1$ using a simple 3-level model.

\section{General relations}
The initial equations of our approach are  very general
equations \cite{r7}
for the one-particle Green functions in a Fermi
system with pairing  G, G$^{h}$, F$^{(1)}$ and F$^{(2)}$
which contain general mass operators $\Sigma$, $\Sigma^{h}$,
$\Sigma^{(1)}$ and $\Sigma^{(2)}$. In order to single out
the well-known components, i.e. the mean field and pairing,
we represent each of the mass operators as a sum of two terms
the first one being energy-independent and the second one energy-dependent:

\begin{eqnarray}
\label{mass}
\Sigma(\varepsilon)=\tilde \Sigma + M(\varepsilon), \hspace{0.2cm}
\Sigma^{(h)}(\varepsilon)=\tilde \Sigma^{(h)} + M^{(h)}(\varepsilon),
\\
\nonumber
\Sigma^{(1,2)}(\varepsilon)=\tilde \Sigma^{(1,2)} + M^{(1,2)}(\varepsilon)=
\tilde \Delta^{(1,2)} + M^{(1,2)}(\varepsilon),
\end{eqnarray}
where
the quantities $\tilde \Sigma$, $\tilde \Sigma^{(h)}$ correspond to
a mean field and $\tilde \Sigma^{(1)}$, $\tilde \Sigma^{(2)}$ describe
a pairing of the BCS type. The quantities M, $M^{(1)}$, $M^{(2)}$, $M^{(h)}$
contain the QPI and do not fix so far.

With taking into account Eqs.~(\ref{mass}) the  general system
for the Green functions
can be transformed into the following equations (see ~\cite{r4,r6} for derivation):
\begin{eqnarray}
\label{solv}
G = \tilde{G} +  \tilde{G}MG - \tilde{F}^{(1)}M^{(h)}F^{(2)} -  \tilde{G}M^{(1)}F^{(2)} - \tilde{F}^{(1)}M^{(2)}G \hspace{1.6cm}
\\
\nonumber
F^{(2)} = \tilde{F}^{(2)} + \tilde{F}^{(2)}MG + \tilde{G}^{(h)}M^{(h)}F^{(2)} - \tilde{F}^{(2)}M^{(1)}F^{(2)} + \tilde{G}^{(h)}M^{(2)}G,
\end{eqnarray}
(and the same for $G^{(h)}$ and $F^{(1)}$).
The bare Green functions
$\tilde{G}$, $\tilde{G}^{(h)}$ and $\tilde F^{(1)}$, $\tilde F^{(2)}$
are the well-known Green functions
 of Gorkov which contain single-particle energies
$\tilde{\varepsilon}_{\lambda}$ and gap $\tilde{\Delta}_{\lambda}$.

In what follows the input parameters of our problem
 are  phenomenological (observed) single-particle energies
$\varepsilon_{\lambda}$ and gap values $\Delta_{\lambda}$
The latter are usually taken from the experiment or obtained
from solving  the BSC equation with a phenomenologically
chosen pp interaction \cite{r2,r3}. Because in the phenomenological
{$\varepsilon_{\lambda},\Delta_{\lambda}$}
there is a contribution
of
 the corresponding M$^{(i)}$ terms,
the quantities
{ $\varepsilon_{\lambda},\Delta_{\lambda}$ }
should be ''refined`` from these contributions to avoid
a double counting of the M$^{(i)}$ terms  or ,in our case, of the QPI.
In other words , a refined basis
{ $\tilde{\varepsilon}_{\lambda},\tilde{\Delta}_{\lambda}$ }
should correspond to the mass operators $\tilde\Sigma^{(i)}$ and
should be used in the  calculations. The procedure to obtain
a connection between
{$\varepsilon_{\lambda},\Delta_{\lambda}$}
and
{ $\tilde{\varepsilon}_{\lambda},\tilde{\Delta}_{\lambda}$ }
has been described in \cite{r4,r6}. The final formulae are the following

\begin{eqnarray}
\label{clear}
 \varepsilon_{\lambda } = \frac{\tilde \varepsilon_{\lambda } + M_{even \lambda}(E_{\lambda})}
{1 + q_{\lambda}(E_{\lambda})}
\\
\nonumber
\Delta_{\lambda} \equiv \Delta_{\lambda}^{(1,2)} = \frac{\tilde \Delta_{\lambda} + M_{\lambda}^{(1,2)}(E_{\lambda})}
{1 + q_{\lambda}(E_{\lambda})}
\end{eqnarray}
where $E_{\lambda} = \sqrt{ \varepsilon_{\lambda}^{2} + \Delta_{\lambda}^{2}}$,
$
\nonumber
q_{\lambda} = - {M_{odd \lambda}(E_{\lambda })}/{E_{\lambda }}
$
and $M_{even}, M_{odd}$ are the even and odd terms of the mass operator $M$.
In equations ~(\ref{clear})
the energies $\tilde \varepsilon_{\lambda}$ and
$\varepsilon_{\lambda}$ are reckoned from the corresponding chemical potential
 $\tilde \mu$ and $\mu$.
These results were obtained by using the approximation which is
diagonal in the single-particle index $\lambda$.

Thus, in order to calculate physical
characteristics 
it is necessary to solve Eqs.~(\ref{clear}) to obtain the new basis
${\tilde{\varepsilon}_{\lambda}, \tilde{\Delta}_{\lambda}}$.
However , from the point of view of the nature of nuclear pairing the
question arises about a possible contribution of the QPI
to the quantity $\tilde{\Delta}_{\lambda}$ because
the difference between $\Delta_{\lambda}$ and $\tilde{\Delta}_{\lambda}$
is only due to the explicit singling out of the QPI contained
in the M$^{(i)}$ terms. Because $\tilde{\Delta}_{\lambda}$
is energy-independent  one can write a general equation for it
(see also~\cite{Acta,r6})
\begin{eqnarray}
\label{new}
\tilde{\Delta}_{\lambda} = \sum_{\lambda'}
W_{\lambda\overline{\lambda}\lambda'\overline{\lambda'}}
 F^{(2)}_{\lambda'\overline{\lambda'}}
\end{eqnarray}
where W is a new energy-independent pp interaction
and F$^{(2)}$ satisfies Eq.(\ref{solv}) in which the quantities
M$^{(i)}$ may be taken in any approximation we need.
Further we will consider our g$^{2}$ approximation
(g$^{2}<1$)  and  a model case of the strong coupling g$^{2}>1$.

\section{g$^{2}$ approximation (g$^{2}<1$)}

In this approximation we should take\\
\begin{eqnarray}
\label{mass0}
\unitlength=1.00mm
\special{em:linewidth 0.4pt}
\linethickness{0.4pt}
\begin{picture}(26.00,6.50)
\put(4.00,0.00){\line(1,0){20.00}}
\put(8.00,1.00){\circle{2.00}}
\put(20.00,1.00){\circle{2.00}}
\put(9.00,3.00){\circle*{1.00}}
\put(10.00,4.00){\circle*{1.0}}
\put(11.00,5.00){\circle*{1.0}}
\put(12.00,5.50){\circle*{1.00}}
\put(13.00,6.00){\circle*{1.00}}
\put(14.00,6.20){\circle*{1.00}}
\put(15.00,6.00){\circle*{1.00}}
\put(16.00,5.50){\circle*{1.00}}
\put(17.00,5.00){\circle*{1.00}}
\put(18.00,4.00){\circle*{1.00}}
\put(19.00,3.00){\circle*{1.00}}
\put(-10.00,1.00){\makebox(0,0)[cc]{\large $\hat{M} =$}}
\end{picture}
\end{eqnarray}
where the circle is the phonon creation amplitude $g$ and
 the single line means the Green function
in its matrix form which includes ${\tilde G}, {\tilde G^{h}}, {\tilde F^{(1)}},{\tilde F^{(2)}}$.
 Here,
the difference between $\Delta_{\lambda}$
and $\tilde{\Delta}_{\lambda}$ is due to the exchange of one
phonon between two quasiparticles (see Eq.(\ref{clear}))
 but, in accordance
with the Eliashberg theory \cite{r1}, this difference
is corrected by the
$(1+q_{\lambda})^{-1}$ factor.

 The system (\ref{clear}) has been solved for the semi-magic nucleus
$^{120}$Sn.  At first phenomenological $\varepsilon_{\lambda}$
and $\Delta_{\lambda}$ were obtained from the existing experimental data
for the  neighbouring $^{119}$Sn and $^{121}$Sn nuclei. The BCS equation
for $\Delta_{\lambda}$ was solved with  the phenomenological volume
pp interaction taken from \cite{r9}.
The phonons have been calculated within the theory of finite
Fermi systems \cite{r2}.We used 21 of the most collective
2$^{+}$, 3$^{-}$, 4$^{+}$, 5$^{-}$ and 6$^{+}$ phonons
with the energy not exceeding the neutron binding energy.
 See \cite{r4},
where values of $\Delta_{\lambda}$ and $\tilde{\Delta}_{\lambda}$
are given, for detailed discussions.
  Here, for simplicity  we use
a simple averaging procedure, for example for $\Delta_{\lambda}$:
\begin{eqnarray}
\Delta_{av}=\frac{\sum_{j}\Delta_{\lambda}(2j+1)}{\sum_{j}(2j+1)}
\end{eqnarray}
    For the  quantity
($\Delta_{av} -\tilde{\Delta}_{av}$)/$\Delta_{av}$,
which gives the QPI contribution caused only by the phonon exchange
between two quasiparticles (or by the retarded pp interaction),
we obtained the value of 31\%.

   Further, we have performed the calculations in the $g^{2}$
approximation for $\tilde{\Delta}_{\lambda}$ according to
Eq.(\ref{new}), i.e. we have solved this equation with the Green
function F$^{(2)}$ obtained from Eq.(\ref{solv}) in our $g^{2}$
approximation  \cite{r6}. The very first term of Eq.(\ref{solv})
is $\tilde{F}^{(2)}$ so that the equation
$\Delta(BCS)=W\tilde{F}^{(2)}$ is a BCS-like equation which
determines a "pure" BCS part of the phenomenological gap value
if the new pp interaction W is known. For simplicity the interaction
W was taken in the same functional form as in \cite{r9}, but the parameter
$c_{p}$ was determined from the condition that the average value
$\tilde{\Delta}_{av}$ found by the solving of the equation obtained
and of the system (\ref{clear}) are the same.
The results of these calculations are as follows.
We obtained
that the contribution of $\Delta_{av}(BCS)$ = 74\% of the average
phenomenological gap , which is 1.42 MeV.
The contribution
of the terms containing the QPI terms
to $\tilde{\Delta}_{av}$ is equal to (-5)\%
of the average
phenomenological gap . This result is similar
to   that for the ph channel in the sense that the contribution
of terms corresponding to phonon exchange diagram and to the self-
energy diagrams are opposite in sign, but in our case the contribution
of the self-energy diagrams in the pp channel is small.

   Thus,  we obtained that the total  contribution of the QPI
to the averaged  phenomenological gap for $^{120}$Sn is
(31-5)=26\% and the BCS-like part is 74\%. One should emphasize
,however, that these are just averaged figures and our method of
determining the new pp interaction is the simplest one.
But in any case we see that pairing in semi-magic nuclei
is of  a mixed nature - it is due to the BCS-like mechanism
and the quasiparticle-phonon one.

   As it was mentioned above, for the calculations of nuclear physical
characteristics
 with taking into account the QPI
it is enough to know the new basis
{$\tilde{\varepsilon}_{\lambda},\tilde{\Delta}_{\lambda}$}.
The question arises
where one can see the effect of this new basis.  The simplest
reply is to compare the calculations with new and old basis.
We have made this comparison (in the framework of our g$^{2}$
approximation) for the spectroscopic factors
in $^{119}$Sn and $^{121}$Sn. The method of calculations
has been described in \cite{r4,r8}. The results are given in
Table 1.
\begin{table}
\begin{center}
\caption{
Spectroscopic factors in $^{119}$Sn (first line)
	       and $^{121}$Sn (second line) calculated with
	       (third column) and without (fourth column)
		taking into account the effect of
 the  new (refined)
		basis.
}
\begin{tabular}{|c|c|c|c|}
\hline
\multicolumn{1}{|c|}{$\lambda$}&
\multicolumn{3}{|c|}{$ S_{\lambda} $}\\
\cline{2-4}
{}&{Exp.~\cite{nds}}&{R+}&{R-}\\
\hline
{$1g7/2$}&{0.75;0.6}&{0.66{\small(0.0-3.0)}}&{0.54}\\
\cline{2-4}
{}&{}&{}&{}\\
\hline
{$2d3/2$}&{0.4;0.45}&{0.4}&{0.28}\\
\cline{2-4}
{}&{0.44;0.65}&{0.35}&{0.42}\\
\hline
{$3s1/2$}&{0.26;0.32}&{0.36}&{0.23}\\
\cline{2-4}
{}&{0.3}&{0.43}&{0.49}\\
\hline
{$1h11/2$}&{0.29}&{0.26}&{0.19}\\
\cline{2-4}
{}&{0.49}&{0.43}&{0.45}\\
\hline
\end{tabular}
\label{clears}
\end{center}
\end{table}

One can see easily that using  our new (refined) basis
improves mainly the description of the experiment. Of course,
it would be desirable to find a better confirmation , i.e.
a clearer manifestation of the effect . Probably
 it can be seen in the calculations of other characteristics
of low-lying levels of odd-mass nuclei. The experiments
using the gamma-spectrometers of the EUROBALL-type can
clarify this question.

\section{Strong coupling (g$^{2}>1$).Model calculations.}

Unfortunately, in this case it is necessary, generally speaking,
to go beyond the Eliashberg-Migdal approximations, i.e.
to add the vertex corrections to the rainbow ones. The reason
is that, because our phonons are made up of the same quasiparticles,
there is no such a small parameter as in the Eliashberg theory
\cite{r1} due to the existence of which these corrections are
negligible. Actually, there can be some specific  reasons
for that in nuclei  (see below) but , in any case, in finite nuclei
we do not have such
a 
small parameter
$g^{2}$
as in the theory of
ordinary superconductors. Thus, instead of the mass operators
(\ref{mass0}),
we should take into account at least
the following mass operators
\begin{eqnarray}
\label{mass1}
\unitlength=1.00mm
\special{em:linewidth 0.4pt}
\linethickness{0.4pt}
\begin{picture}(80.00,6.50)
\put(9.00,1.00){\line(1,0){10.00}}
\put(4.00,0.00){\line(1,0){20.00}}
\put(8.00,1.00){\circle{2.00}}
\put(20.00,1.00){\circle{2.00}}
\put(9.00,3.00){\circle*{1.00}}
\put(10.00,4.00){\circle*{1.0}}
\put(11.00,5.00){\circle*{1.0}}
\put(12.00,5.50){\circle*{1.00}}
\put(13.00,6.00){\circle*{1.00}}
\put(14.00,6.20){\circle*{1.00}}
\put(15.00,6.00){\circle*{1.00}}
\put(16.00,5.50){\circle*{1.00}}
\put(17.00,5.00){\circle*{1.00}}
\put(18.00,4.00){\circle*{1.00}}
\put(19.00,3.00){\circle*{1.00}}
\put(33.00,1.00){\line(1,0){4.00}}
\put(28.00,0.00){\line(1,0){20.00}}
\put(32.00,1.00){\circle{2.00}}
\put(44.00,1.00){\circle{2.00}}
\put(33.00,3.00){\circle*{1.00}}
\put(34.00,4.00){\circle*{1.0}}
\put(35.00,5.00){\circle*{1.0}}
\put(36.00,5.50){\circle*{1.00}}
\put(37.00,6.00){\circle*{1.00}}
\put(38.00,6.20){\circle*{1.00}}
\put(39.00,6.00){\circle*{1.00}}
\put(40.00,5.50){\circle*{1.00}}
\put(41.00,5.00){\circle*{1.00}}
\put(42.00,4.00){\circle*{1.00}}
\put(43.00,3.00){\circle*{1.00}}
\put(34.00,0.00){\line(1,0){20.00}}
\put(38.00,1.00){\circle{2.00}}
\put(50.00,1.00){\circle{2.00}}
\put(39.00,3.00){\circle*{1.00}}
\put(40.00,4.00){\circle*{1.0}}
\put(41.00,5.00){\circle*{1.0}}
\put(42.00,5.50){\circle*{1.00}}
\put(43.00,6.00){\circle*{1.00}}
\put(44.00,6.20){\circle*{1.00}}
\put(45.00,6.00){\circle*{1.00}}
\put(46.00,5.50){\circle*{1.00}}
\put(47.00,5.00){\circle*{1.00}}
\put(48.00,4.00){\circle*{1.00}}
\put(49.00,3.00){\circle*{1.00}}
\put(-10.00,1.00){\makebox(0,0)[cc]{\large $\hat{M} =$}}
\put(26.0,1.0){\makebox(0,0)[cc]{\large +}}
\end{picture}
\end{eqnarray}
where the double line means the full Green function
in its matrix form which includes G, G$^{h}$, F$^{(1)}$, F$^{(2)}$.

It is necessary now to solve Eqs.~\ref{solv}
 with these mass operators.
For the qualitative understanding and analysis of the case with strong coupling, here
we consider  a schematic 3-level model. We will take
three levels with energies $\varepsilon$=-1, 0, 1, each
of them has the quantum number $j=11/2$, the first one being
occupied completely, parity of the low- and high-lying levels
being positive and that of intermediate level negative.
The occupation number of the intermediate level will
be changed from 0 to $2j+1=12$. In this model , to obtain
$\Delta \ne$ 0, the parameter of pp interaction should be
$G_{pp}$=0.035. The ph interaction
was taken in our case  to obtain the Bohr parameter
$\alpha = g^{2}/((2j+1)\omega_{s})$ ($\omega_{s}$- is the energy of phonon)
 equal to 2 for
the most
collective $2^{+}$ phonon.
\begin{figure}[t]
\centerline{\includegraphics[width=0.8\linewidth,height=1.0\linewidth,angle=-90]{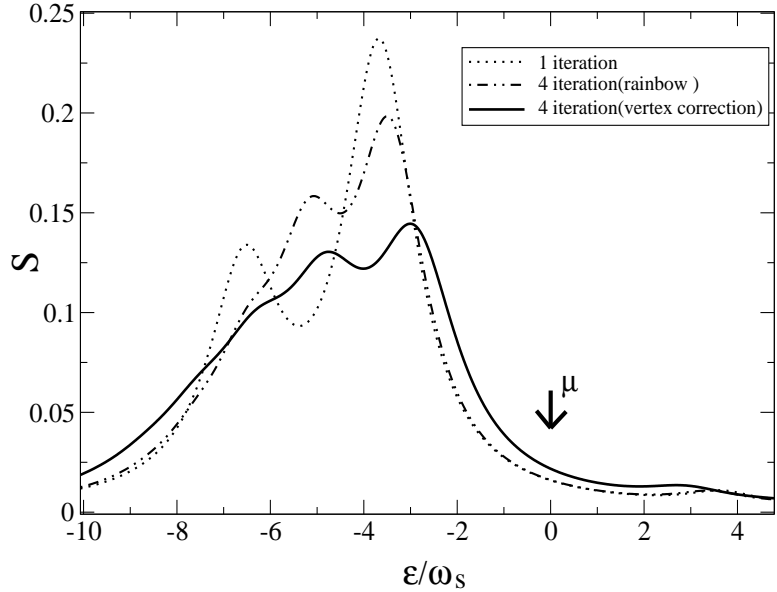}}
\caption{
 Distribution of the single-particle strength calculated
in the three-level model for the case with pairing, j=11/2,
$\alpha$=2 ( see the text for details)
The dispersion
was taken as a
  smearing parameter for each of the  calculations.
}
\end{figure}

To estimate the contribution of the vertex correction
we used the well- known results
by Belyaev and Zelevinsky \cite{BZ} according to which
every vertex contains the factor $w$ with a 6j-symbol
$
\omega=-(2j+1)\left\{
\begin{array}{rcl}
j&2&j\\
j&2&j\\
\end{array}
\right\}
$
so that the n-iteration contains a factor $w^{n}$ which
decreases the contribution of the vertex correction as compared
to the rainbow contribution. In our case $w$=0.76, so
in what follows we restricted ouselves to $w$ only. That means
that at least the g$^{4}$ terms are taken into account
reasonably enough
in our nonlinear model with the mass operators (\ref{mass1}).
For the beginning and simplification we have calculated
the spectroscopic factors in our non-linear model
with the mass operators (\ref{mass1}) , the above-mentioned
parameter $\alpha$  being taken $\alpha$=2, which corresponds
to a realistic nuclear case.
To obtain the distribution of the single-particle strength
the iteration procedure was organized for the solution of
the Dyson equation Eq. (\ref{solv}) with our mass operators.
In our case  this procedure must include also the refinement
procedure for $\tilde{\varepsilon}_{i}$ and $\tilde{\Delta}_{i}$.

The results are given in Fig.1. It turned out that it was necessary
to do four iterations to obtain the convergency. This corresponds
to the approximation 1quasiparticle$\otimes$ 4 phonons.
  We can see that the inclusion of the vertex correction
is noticeable in the strength distribution.

 In particular,
 for the dominant level we obtained the decrease of the
spectroscopic factor from 0.49
to 0.40 due to the inclusion of the vertex corrections.
The calculations have shown also that the role of the mass
operators M$^{(1)}$ and M$^{(2)}$ for the levels far off
the Fermi surface was small.

\section{Conclusion}
We obtained that,
if the simple
procedures proposed are used to determine the new particle- particle
interaction and to estimate the average effect according to
Eq.(6), then
the QPI contribution was 26\%  and the BCS-like mechanism gave 74\%
of the gap observed for $^{120}$Sn. This means that at least
for semi-magic nuclei pairing is of a mixed nature-it is
due to the quasiparticle-phonon and the BCS-like
mechanisms, the first being mainly a surface one and the second
mainly a volume mechanism. The effect of the  QPI in the gap value
can be observed probably
in the experiments
using  the gamma-spectrometers of the
EUROBALL-type to measure the characteristics of low-lying levels.

For the case of nuclei with unfilled shells in both proton
and neutron systems (strong coupling) our estimation has shown that
the contribution of the vertex correction to spectroscopic
factors is rather noticiable. One can think that this
contribution is
even
 more noticeable in
 transition probabilities for  low-lying levels.

 S.K. thanks Prof. G.M. Eliashberg for useful
discussions of the results.

\end{document}